\documentclass[letterpaper,twocolumn,10pt]{article}
\usepackage{usenix2019_v3}
\usepackage{epsfig,endnotes, graphicx, tikz, pgfplots, csvsimple, array, multirow, subfig, caption}
\pgfplotsset{width=7cm,compat=1.9}
\usetikzlibrary{calc,trees,positioning,arrows,chains,shapes.geometric,%
    decorations.pathreplacing,decorations.pathmorphing,shapes,%
    matrix,shapes.symbols}

\usepackage[skip=3pt,font=small]{caption}
\usepackage{fancyhdr}

\usepackage{pgf}
\usepackage{tikz}
\usetikzlibrary{arrows,automata, positioning}
\usepackage{hyperref}

\begin{document}

\date{}

\title{\Large \bf FlexServe: Deployment of PyTorch Models as Flexible REST Endpoints}

\author{
{\rm Edward Verenich$^{1,2}$}, {\rm Alvaro Velasquez$^{1}$}, {\rm M.G. Sarwar Murshed$^{2}$}, {\rm Faraz Hussain$^{2}$}\\
$^{1}$Air Force Research Laboratory, $^{2}$Clarkson University
}

\maketitle


\subsection*{Abstract}
The integration of artificial intelligence capabilities into modern software systems
is increasingly being simplified through the use of cloud-based machine learning services and representational state transfer architecture design.
However, insufficient information regarding underlying model provenance and
the lack of control over model evolution serve as an impediment to the
more widespread adoption of  these services in many operational environments
which have strict security requirements.
Furthermore, tools such as TensorFlow Serving allow models to be deployed as RESTful endpoints, but require error-prone transformations for PyTorch models as these dynamic computational graphs. This is in contrast to the static computational graphs of TensorFlow.
To enable rapid deployments of PyTorch models without intermediate transformations we have developed FlexServe, a simple library to deploy multi-model ensembles with flexible batching.

\section{Introduction}\label{sec:intro}

The use of machine learning (ML) capabilities, such as visual inference and classification, in operational software systems continues to increase. A common approach for incorporating ML functionality into a larger operational system is to isolate it behind a micro service accessible through a REpresentational State Transfer (REST) protocol~\cite{10.5555/2969442.2969519}.
This architecture separates the complexity of ML components from the rest of the application while making the capabilities more accessible to software developers through well-defined interfaces. 

Multiple commercial vendors offer such capabilities as cloud services as described by Cummaudo et al in their assessment of using such services in larger
software systems~\cite{DBLP:journals/corr/abs-1906-07328}.
They found a lack of consistency across services for the same data points,
and also the unpredictable evolution of models,
resulting in temporal inconsistency of a given service for identical data points.
This behavior is due to the underlying classification models and their evolution as they are trained on different and additional data points by their vendors, providing insufficient information to the consuming system developer regarding the provenance of the model. Lack of control over the underlying models prevents many operational systems from consuming inference output from these services.

A better approach for preserving the benefits of a REST architecture while maintaining control of all aspects of model behavior is to deploy them as RESTful endpoints, thereby exposing them to the rest of the system.
A popular approach for serving ML models as REST services is TensorFlow Serving~\cite{olston2017tensorflow}. However, serving PyTorch's dynamic graph through Tensor Flow Serving is not possible
and requires transforming PyTorch models to an intermediate representation such as Open Neural Network Exchange (ONNX) format~\cite{bai2019} which in turn is converted to the TensorFlow static graph compatible with TensorFlow Serving.
This two-step conversion often fails and is difficult to debug, because not all PyTorch features are supported by ONNX,
making the train-test-deploy pipeline through TensorFlow Serving at best slow and at worst impossible for some PyTorch models. Another solution is to use a Kubernetes library KFServing \cite{KFServing}, which is currently in beta status, but it requires the deployment of Kubernetes, as KFServing is a serverless library and depends on a separate ingress gateway deployed on Kubernetes. Although KFServing is a good solution if Kubernetes is available, its deployment options are not lightweight when compared to TensorFlow Serving and Kubernetes itself is complex to configure and manage \cite{Kubernetes}.

\begin{figure*}[t]
    \centering
    \includegraphics[width=16cm]{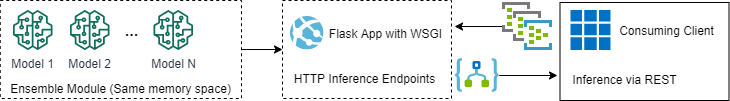}
    \caption{FlexServe architecture consists of an ensemble module which loads N of models into a shared memory space. Inference output of each model is combined in a single response and returned to the requesting client as a JSON response object. The Flask application invokes the \emph{fmodels} module, which encompasses the ensemble of models, and exposes RESTful endpoints through the Web Service Gateway Interface. Variable batch sizes provide for maximum efficiency and flexibility as clients are not restricted to a fixed batch size of samples to send to the inference service. Additional efficiency is achieved through the use of the same memory space, better utilizing available GPUs and  requiring only one data transformation for all models in the ensemble.}
    \label{fig:flexserve}
\end{figure*}

To enable PyTorch model deployments in a manner similar to TensorFlow model deployments with TensorFlow Serving, \emph{we  have developed FlexServe, a simple REST service deployment module} that provides the following additional capabilities which are commonly needed in operational environments: (i) the deployment of multiple models behind a single endpoint, (ii) the ability to share a single GPU memory across multiple models, and (iii) the capacity to perform inference using flexible batch sizes.

In the remainder of the paper, we describe FlexServe and demonstrate its advantages in scenarios having requirements outlined above.

\section{Approach and Use Cases}

FlexServe is implemented as a Python module encompassed in a Flask \cite{Flask} application. Flask is a web application framework that facilitates web application development and communication across web services. We chose the lightweight Gunicorn \cite{Gunicorn} application server as the Web-Server Gateway Interface (WSGI) HTTP server to be used within the Flask application. This WSGI enables us to forward requests to web applications using the Python programming language.  Figure~\ref{fig:flexserve} shows the high-level FlexServe architecture and its interaction with consuming applications.

\subsection{Multiple models, single endpoint}
Running ensembles of models is a common way to improve classification accuracy~\cite{7780459},
but it can also be used to adjust sensitivity (number of false negatives) of the ensemble dynamically. Consider an ensemble of $n$ models trained to recognize the presence of a specific object. By using different architectures, the ensemble model takes advantage of different inductive biases that perform better at different geometric variations of the target object. Then, combining its inference outputs according to the sensitivity policy of the consuming application, ensemble sensitivity can be adjusted dynamically. For example, let $y\in \{0,1\}$ be a binary output (0=absent, 1=present) of a classifier and let $y'$ be the combined output. Then for maximum sensitivity the policy is $y'= y_1 | y_2 | ...| y_n$, meaning that when a single model detects the target the final ensemble output is positive identification. Different sensitivity policies can be employed by the client as needed.
\subsection{Share a single device}
Deployed models vary in size, but are often significantly smaller than the memory available on hardware accelerators such as GPUs. Loading multiple models in the same device memory brings down the cost and provides for more efficient inference. FlexServe allows multiple models to be loaded as part of the ensemble and performs multi-model inference on a single \emph{forward} call of the \emph{nn.Module}, thereby removing the additional data transformation calls associated with competing methods. Scaling horizontally to multiple CPU cores is also possible through the use of Gunicorn workers.
\subsection{Varying batch size}
FlexServe accepts varying batch sizes of image samples and returns a combined result of the form \textit{'model y\_i': ['class','class',...,'class']} for every model $y_i$.
This functionality can be used in many applications. For example, to perform time series tracking from conventional image sensors or inexpensive web cameras by taking images at various time intervals and sending these chronological batches of images to FlexServe. As an object moves through the field of view of the sensor, a series of images is produced that can be used to infer movement of an object through the surveillance sector when more sophisticated object trackers are not available and video feeds are too costly to transmit. This places the computation and power burden on the Flask server as opposed to the potentially energy-constrained consumer which is only interested in the inference results of the ensemble model.

\section{Conclusion}\label{sec:conclusion}
Commercial cloud services offer a convenient way of introducing ML capabilities to software systems through the use of a representational state transfer architecture. However, lack of control over underlying models and insufficient information regarding model provenance and evolution limit their use in many operational environments.
Existing solutions for deploying models as RESTful services
are not robust enough to work with PyTorch models in many environments.
FlexServe is a simple and lightweight solution that provides deployment functionality similar to TensorFlow Serving without intermediate conversions to a static computational graph.

\section*{Availability}
The FlexServe deployment module and example are available
 in a public repository: {\tt \url{https://github.com/verenie/flexserve}}

\section*{Acknowledgments}
This authors acknowledge support from the StreamlinedML program (Notice ID FA875017S7006) of the Air Force Research
Laboratory Information Directorate.

{\footnotesize \bibliographystyle{unsrt}
\bibliography{refs}}

\begin{thebibliography}{1}

\bibitem{10.5555/2969442.2969519}
D.~Sculley, Gary Holt, Daniel Golovin, Eugene Davydov, Todd Phillips, Dietmar
  Ebner, Vinay Chaudhary, Michael Young, Jean-Francois Crespo, and Dan
  Dennison.
\newblock Hidden technical debt in machine learning systems.
\newblock In {\em Proceedings of the 28th International Conference on Neural
  Information Processing Systems - Volume 2}, NIPS’15, page 2503–2511,
  Cambridge, MA, USA, 2015. MIT Press.

\bibitem{DBLP:journals/corr/abs-1906-07328}
Alex Cummaudo, Rajesh Vasa, John~C. Grundy, Mohamed Abdelrazek, and Andrew
  Cain.
\newblock Losing confidence in quality: Unspoken evolution of computer vision
  services.
\newblock {\em CoRR}, abs/1906.07328, 2019.

\bibitem{olston2017tensorflow}
Christopher Olston, Noah Fiedel, Kiril Gorovoy, Jeremiah Harmsen, Li~Lao,
  Fangwei Li, Vinu Rajashekhar, Sukriti Ramesh, and Jordan Soyke.
\newblock Tensorflow-serving: Flexible, high-performance ml serving.
\newblock {\em arXiv preprint arXiv:1712.06139}, 2017.

\bibitem{bai2019}
Junjie Bai, Fang Lu, Ke~Zhang, et~al.
\newblock Onnx: Open neural network exchange.
\newblock \url{https://github.com/onnx/onnx}, 2019.

\bibitem{KFServing}
Kfserving.
\newblock \url{https://github.com/kubeflow/kfserving}.
\newblock Accessed: 2020-02-25.

\bibitem{Kubernetes}
Kelsey Hightower, Brendan Burns, and Joe Beda.
\newblock {\em Kubernetes: up and running: dive into the future of
  infrastructure}.
\newblock " O'Reilly Media, Inc.", 2017.

\bibitem{Flask}
Flask.
\newblock \url{https://github.com/pallets/flask}.
\newblock Accessed: 2020-02-25.

\bibitem{Gunicorn}
Gunicorn.
\newblock \url{https://gunicorn.org/}.
\newblock Accessed: 2020-02-25.

\bibitem{7780459}
K.~{He}, X.~{Zhang}, S.~{Ren}, and J.~{Sun}.
\newblock Deep residual learning for image recognition.
\newblock In {\em 2016 IEEE Conference on Computer Vision and Pattern
  Recognition (CVPR)}, pages 770--778, June 2016.

\end{thebibliography}


\end{document}